\documentclass[12pt]{aastex}
\usepackage{emulateapj5}

\newcommand{\um}{$\mu$m}

\newcommand{\MSX}{{\it MSX}}
\newcommand{\iso}{{\it ISO}}

\newcommand{\msx}{{\it MSX}}

\newcommand{\degree}{$^{\circ}$}

\newcommand{\drco}{$^{13}{\rm CO}$}

\newcommand{\myemail}{simonr@ph1.uni-koeln.de, jackson@bu-ast.bu.edu, rathborn@bu.edu, etc1@bu.edu}
\newcommand{\mum}{$\mu$m}
\newcommand{\jeins}{J=1$\to$0}

\newcommand{\msxunit}{W m$^{-2}$ sr$^{-1}$}

\slugcomment{Version of \today}

\shortauthors{Simon et al.}
\shorttitle{\MSX\, Infrared Dark Cloud Catalog}

\begin{document}

\title{A Catalog of \MSX\, Infrared Dark Cloud Candidates}

\author{Robert Simon\altaffilmark{1}, James M. Jackson, Jill M. Rathborne, and
  Edward T. Chambers} 
\affil{Institute for Astrophysical Research, Boston University, 725 Commonwealth
  Avenue, Boston, MA 02215} \altaffiltext{1}{Current address: I. Physikal.
  Institut, Universit\"at zu K\"oln, Z\"ulpicher Str. 77, 50937 K\"oln, Germany}
\email{\myemail}

\begin{abstract}
  We use 8.3\,\um\ mid-infrared images acquired with the {\it Midcourse Space
    Experiment} satellite to identify and catalog Infrared Dark Clouds (IRDCs)
  in the first and fourth quadrants of the Galactic plane. Because IRDCs are
  seen as dark extinction features against the diffuse Galactic infrared
  background, we identify them by first determining a model background from the
  8.3 \um\ images and then searching for regions of high decremental contrast
  with respect to this background. IRDC candidates in our catalog are defined by
  contiguous regions bounded by closed contours of a $ 2 \sigma$ decremental
  contrast threshold.  Although most of the identified IRDCs are actual cold,
  dark clouds, some as yet unknown fraction may be spurious identifications. 
  For large, high contrast clouds, we estimate the reliability to be 82\%.
  Low contrast clouds should have lower reliabilities. Verification of the reality of individual clouds will
  require additional data.  We identify 10,931 candidate infrared dark clouds. For each
  IRDC, we also catalog cores.  These cores, defined as
  localized regions with at least 40\% higher extinction than the cloud's
  average extinction, are found by iteratively fitting 2-dimensional elliptical
  Gaussians to the contrast peaks.  We identify 12,774 cores.  The catalog
  contains the position, angular size, orientation, area, peak contrast, peak 
contrast signal-to-noise, and integrated contrast of the candidate IRDCs and 
their cores. The distribution of IRDCs
  closely follows the Galactic diffuse mid-infrared background and peaks toward
  prominent star forming regions, spiral arm tangents, and the so-called 5\,kpc 
  Galactic molecular ring.
\end{abstract}

\keywords{catalogs ---dust, extinction ---Galaxy: general ---infrared: ISM ---
  ISM: clouds ---ISM: structure}

\section{Introduction}

Since the early identifications of optically dark regions
\citep[e.g.,][]{Barnard1919,Bok1947,Lynds1962,Feitzinger1984}, extinction
apparent from stellar density fluctuations or against bright, diffuse background
emission has been used to identify, catalog, and study interstellar clouds. In
recent years, such studies have started to exploit increasingly higher angular
resolutions and longer wavelengths. Since extinction is much smaller at infrared
(IR) wavelengths compared to the optical, the distances out to which dark clouds can
be detected, as well as the dynamic range in magnitudes of extinction, have
become much larger.  With the advent of large scale, multi-color IR
surveys, such as the Two Micron All Sky Survey (2MASS; \citealp{Skrutskie1997}) and 
the {\it Spitzer} Galactic Legacy Infrared Mid-Plane Survey Extraordinaire (GLIMPSE; \citealp{Benjamin03}), 
extinction studies have become even more
sophisticated, combining number density fluctuations of stars with color 
information \citep[e.g.,][]{Lada94,Cambresy02}.

Recent high resolution, high sensitivity IR surveys of the Galactic
plane, performed with the {\it Infrared Space Observatory} (\iso; \citealp{Perault96})
and {\it Midcourse Space Experiment} (\msx; \citealp{Egan98}) satellites 
have identified a new class of interstellar
clouds, the Infrared Dark Clouds (IRDCs). IRDCs are clouds of high extinction
($>2$ mag at 8 \mum; \citealp{Egan98}) seen in silhouette against the bright
Galactic background at mid-IR wavelengths \citep{Carey98,Hennebelle01}.

Preliminary millimeter and submillimeter studies of a few IRDCs show they are
dense ($>10^{5}$ cm$^{-3}$), cold ($<25$~K), and have very high column densities
($\sim 10^{23}$--$10^{25}$~cm$^{-2}$; \citealp{Egan98,Carey98,Carey00}). IRDCs
are often completely opaque at wavelengths between 7 and 100 \um\ 
\citep{Carey98,Teyssier02}.  The low temperatures of IRDCs indicate that
they lack the massive young stars that can be associated with dense molecular gas
clouds. These properties make them excellent candidates to be in the
earliest stages of star formation. Indeed, some evidence suggests
ongoing, deeply embedded star formation in a few IRDCs
\citep{Redman03,Ormel05}. Very recently, three massive protostars toward an
IRDC have also been detected \citep{Rathborne05}.

Despite their potential importance, little is known about the origin and
Galactic distribution of IRDCs. In order to understand the role of IRDCs in star
formation and the Galactic interstellar medium, it is important to identify and
catalog a large, uniform sample. To produce such a catalog, a sensitive, high
angular resolution mid-IR survey is necessary. The \MSX\, Galactic Plane Survey
\citep{Price01} provides the ideal dataset. Launched in 1996, the {\it MSX}
SPIRIT III 33.7 cm telescope was used to survey the Galactic Plane at four
mid-IR wavelengths centered on 8.3, 12.1, 14.7, and 21.3 \mum\ (Bands A, C, D,
and E, respectively). The 8.3\,$\mu$m (Band A) data offer the best combination
of angular resolution and sensitivity (20\arcsec, 1.3~MJy sr$^{-1}$;
\citealp{Price01}), and are far superior to similar images obtained with {\it
  IRAS} (4\arcmin\,$\times$ 5\arcmin, 0.05~MJy sr$^{-1}$ at 12 \um).  In addition,
because the \msx\ 8.3 \um\ filter contains emission from PAH molecules at 7.7
and 8.6 \mum, it has the brightest diffuse mid-IR background emission from the Galactic plane. IRDCs
will thus be most easily detectable in this band.

An \MSX\, IRDC catalog will be a valuable resource for finding candidate star
forming clouds and cores in their early stages of evolution.  
However, because of the inherent difficulties in identifying extended structures,
it must be understood that this catalog contains only {\it candidate} cores.
It will, however, help to support a number of existing and planned IR surveys. For example, the
study of IRDCs will form a key component of the GLIMPSE {\it Spitzer} Legacy project.
Identification of a large sample of IRDCs will also allow follow-up, higher
angular resolution studies by future space-based missions such as Herschel and
the JWST, the airborne SOFIA telescope, and large ground-based IR telescopes
such as the IRTF, Gemini, and Subaru. The {\it MSX} IRDC Catalog will provide
the finding charts by which to characterize IRDCs and their embedded cores
throughout the Galaxy.

\section{Generating the IRDC Candidate Catalog}
\label{irdcfinding}

\begin{figure*}
\epsscale{.80}
\plotone{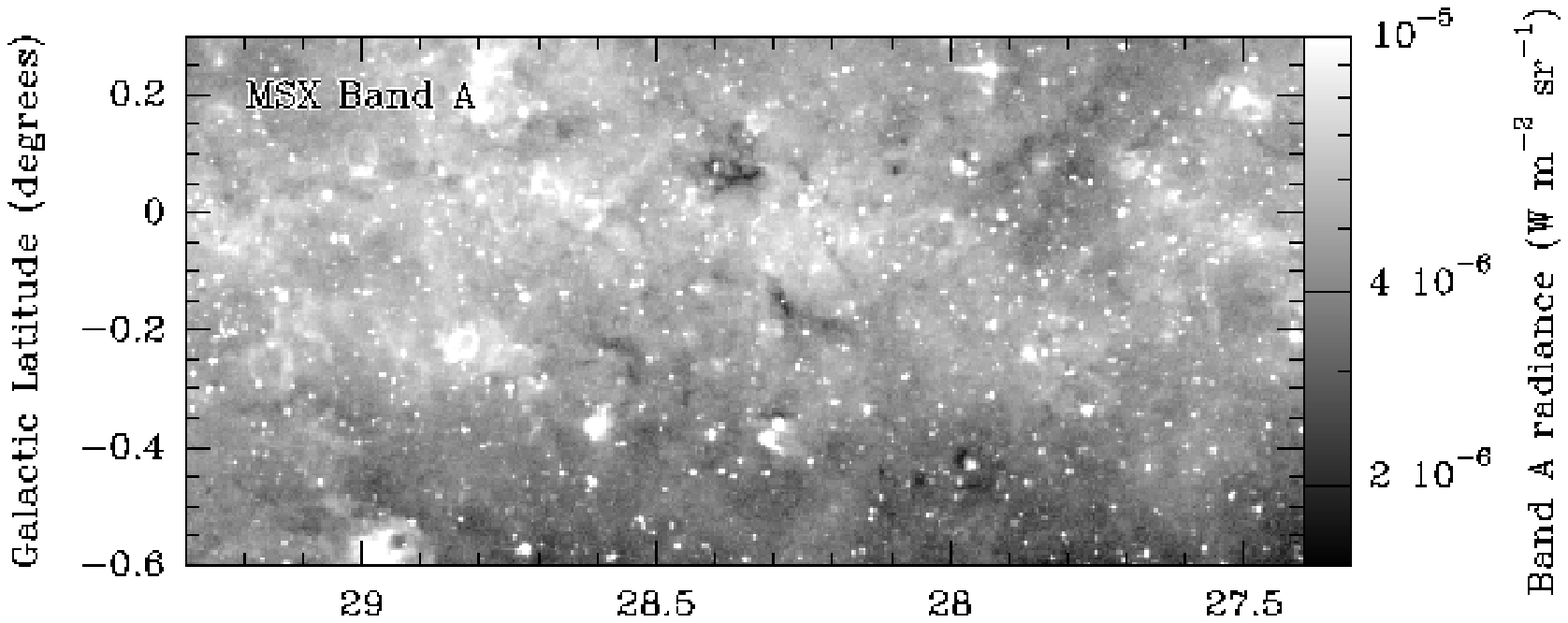}
\plotone{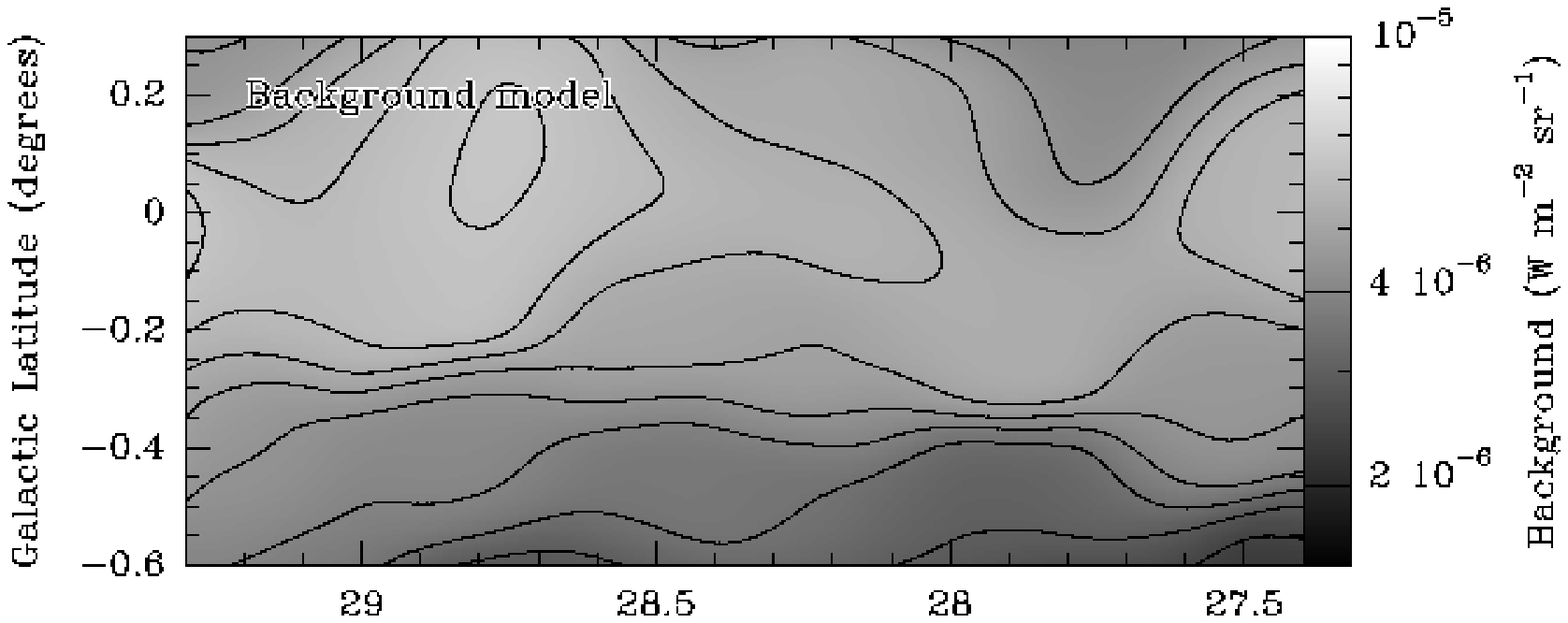}
\plotone{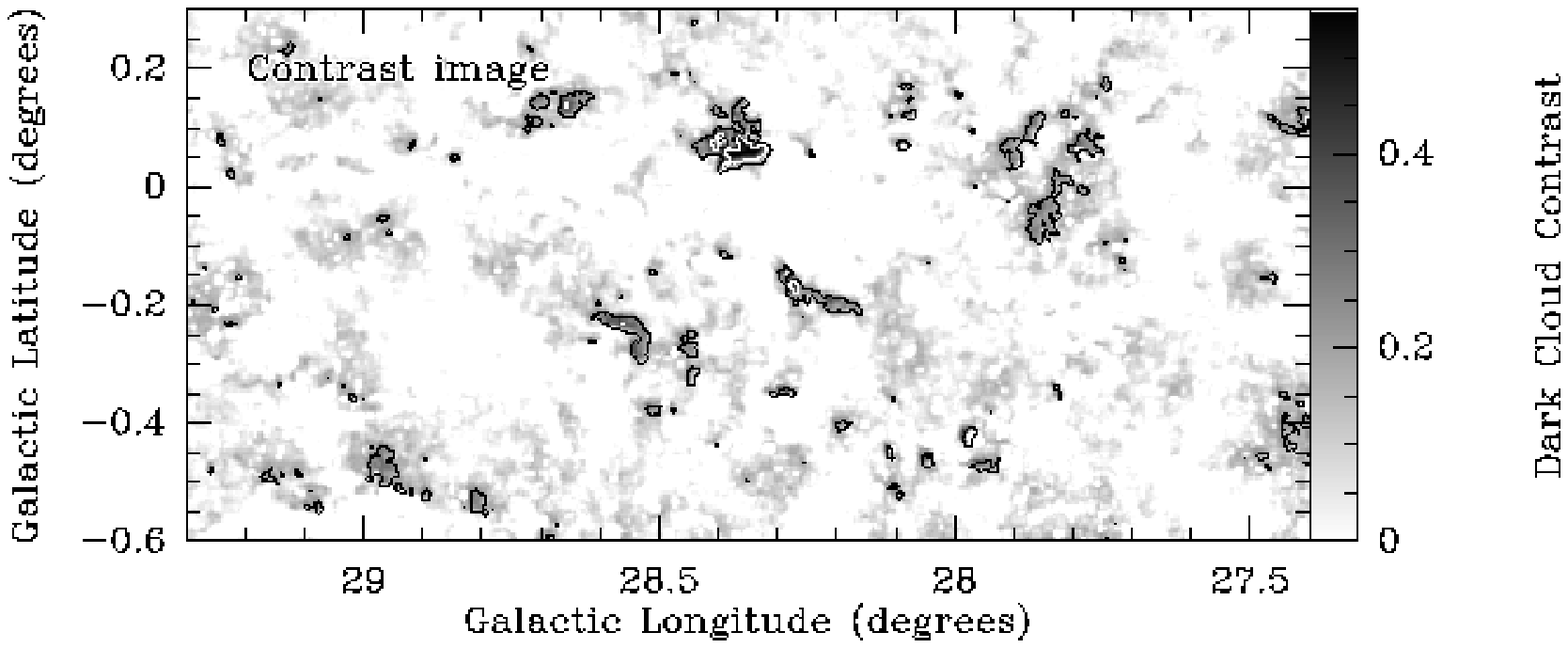}
\caption{Top: \msx\ 8.3 \um\ (Band A) image of a region along the Galactic plane around 
  $l=28$\degree. The intensity scale is logarithmic from $1.5 \times 10^{-6}$
  (black) to $1.0 \times 10^{-5}$ W m$^{-2}$ sr$^{-1}$ (white). Middle:
  Background model.  Contours are from $2.5\times 10^{-6}$ to $6.5\times
  10^{-6}$ W m$^{-2}$ sr$^{-1}$ in steps of $0.5\times 10^{-6}$ W m$^{-2}$
  sr$^{-1}$.  Bottom: Contrast image determined from the original data and the
  background model. Contours are plotted for contrast values of 0.2 (black) and
  0.4 (white).}
\label{fig1}
\end{figure*}

In this section, we outline our method for producing the catalog of IRDCs from
the 8.3\,\um\, \MSX\, images.  Because IRDCs are revealed as extinction
features, we define IRDCs as extended regions that show a significant decrement
in the mid-IR background.  In order to identify and catalog IRDCs, we have
written an automated algorithm that selects IRDCs from the \msx\ Galactic Plane
Survey. We used the \msx\ Band A plates available at the NASA/IPAC Infrared
Science Archive (IRSA\footnote{IRSA is operated by the Jet Propulsion
  Laboratory, California Institute of Technology, under contract with the
  National Aeronautics and Space Administration. See http://irsa.ipac.caltech.edu.}). 
Each image is a Cartesian
projection about the Galactic Center and has 6\arcsec\ pixels and 20\arcsec\ 
angular resolution.  From the original \msx\ images (1.5 by 1.5 degrees in
size), we generated larger mosaics of 4.5 degree width in Galactic longitude and
the full height of the \msx\ coverage in Galactic latitude ($\pm 5$ degrees).
Adjacent mosaics overlap by 1.5 degrees in longitude. Our IRDC identification
algorithm contains three major steps: (1) modelling the diffuse IR background,
(2) generating an image of decremental contrast against the background, and (3)
identifying and selecting candidate IRDCs.  The overall methodology described here is also
illustrated in Fig.~\ref{fig1}. Each of these steps is described in detail in
the following sections.

\subsection{Modelling the Diffuse IR Background}

We assume that the 8.3\,\um\, images, apart from bright emission from small
regions or point sources, consist of a diffuse, extended IR background against
which the IRDCs are superposed. To model this background, we must retain only
the slowly varying spatial components of the images.  We have chosen the
technique of spatial median filtering to generate our model background.

For each pixel, the diffuse background intensity is estimated by selecting the
median value of the intensity distribution of all pixels inside a circular
region centered on each pixel.  The size of the circular region must be larger
than the typical size of IRDCs; otherwise, the decrement due to the presence of
an IRDC will be considered a fluctuation in the background.  On the other hand,
it must also be small enough to follow the real spatial fluctuations in the
background. 

Because typical sizes of IRDCs are smaller than $10'$ \citep{Carey00},
the filter should be larger than $\sim 10'$. On the other
hand, because typical spatial variations in the Galactic plane 8 \um\,
background are typically larger than $\sim 30'$ (estimated from cuts through the
original \msx\ data parallel and perpendicular to the plane at various
longitudes), the filter should also be no larger than $\sim 30'$. With these
considerations in mind, we varied the filter size through a wide range of
values, and finally selected a circular region with a radius of $15'$ (150
pixels for the original 6\arcsec\ \MSX\, pixel size).

\begin{figure}[h,t]
\epsscale{1.0}
\plotone{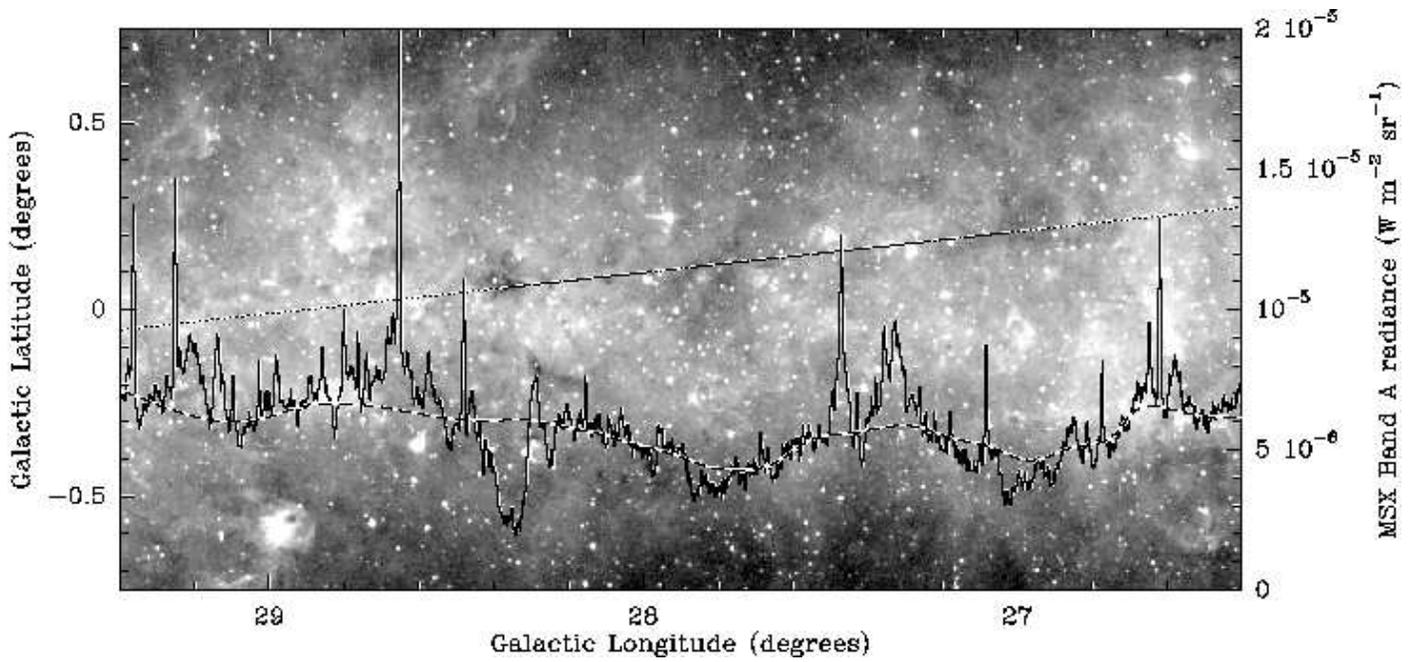}
\caption{
  Distribution of \msx\ 8.3 \um\, (Band A) intensities in a cut along the
  Galactic plane around $l=28$\degree\ (histogram, intensity axis on the right)
  overlaid on the \msx\ 8.3 \um\, image.  The direction of the cut is indicated
  by the dotted line crossing the image. The dashed line shows the background
  intensities, calculated as described in the text, along the same cut. Several
  intensity decrements relative to the background corresponding to IRDCs 
  are seen in the image and the intensity histograms. The most prominent
  dark cloud is at $l=28$\fdg35.  The intensity scale in the image is $1.5
  \times 10^{-6}$ (black) to $1.0 \times 10^{-5}$ W m$^{-2}$ sr$^{-1}$ (white)
  on a logarithmic scale.}
\label{fig2}
\end{figure}

We tested the reliability of our background model by inspecting cuts in various
directions over a wide range of longitudes. In the majority of cases for our
choice of a $15'$ radius circular spatial filter, the modelled background
intensity very closely follows the variation in the \MSX\ images and appears to
yield a satisfactory background estimate. Fig.~\ref{fig1} shows an original
\MSX\, 8.3 \um\ image and the corresponding model background for a typical
field. In addition, Fig.~\ref{fig2} shows a plot of the background model
superposed on the original image data for a cut through a larger field.

\subsection{The Contrast Image}

Once the diffuse background is estimated, we then generate an image of the
decremental contrast against that background.  In order to enhance brightness
sensitivity and suppress spurious detections in the noisier (relative to the
background) low background, higher latitude regions, we first convolved the
\msx\ images with a 20\arcsec\ Gaussian.  The resulting images thus have an
effective angular resolution of 28\arcsec. The smoothed 8.3 \um\, image is then
subtracted from the diffuse background image, and then divided by the background
image, to produce the `contrast' image: Contrast=(Background--Image)/Background.
With this definition, regions that reveal a decrement against the diffuse
background will have a positive contrast.  The lower panel of Fig.~\ref{fig1}
shows a contrast image calculated from the data and background (top and middle
panel). All regions showing obvious extinction in the
\msx\ image are clearly identified as high contrast features, in particular the
IR dark cloud at $l=28$\fdg35, whose contrast of $\sim 0.6$ is among the highest
values found in the Galactic plane.

\subsection{IRDC Candidate Identification}

We then examine the contrast image to select IRDCs.  We require that IRDCs be
isolated from one another, continuous, and extended.  Using a method of
closed-contour identification, we identify IRDCs as regions which contain
contiguous pixels above a certain contrast threshold and angular size (using
algorithms developed using the data reduction package {\sc
  graphic}\footnote{Part of the {\sc gildas} software package maintained by
  IRAM/Grenoble}).

All IRDC candidates must meet the following two criteria to be included in the catalog. First, the
contrast values for the contiguous regions must be large enough to be real, and
not the result of instrumental noise. The uncertainty in the contrast for each
pixel was calculated using the rms noise fluctuations in the \msx\ images in
regions without obvious emission away from the plane ($2.1 \times 10^{-7}$
\msxunit\ for the smoothed, 28\arcsec\ resolution, images). We identify as
candidate IRDCs only those regions whose contrasts are at least twice the error
($2\sigma$) estimated for each of the contiguous pixels.  Second, the IRDCs must
be extended in order to avoid identification of very small features, which may
be artefacts due to noisy pixels in the \msx\ images.  Consequently, we require
that IRDCs have an equivalent area, defined by the area enclosed by 10\% of the
peak contrast level, to be larger than twice the convolved 28\arcsec\ FWHM
beam solid angle, $\Omega > {{\pi}\over{2}} (28 \arcsec)^2 = 1,232$ square
arcsec.  In practice, we retain only those sources with 36 or more contiguous
pixels (1,296 square arcsec). 
  
The 10,931 sources thus identified comprise the candidate clouds in our {\it MSX} IRDC
catalog.  Fig.~\ref{clouds+cores} shows several of the clouds (marked as black
ellipses) identified toward a subset of the region in the Galactic plane shown
in Fig.~\ref{fig1} and \ref{fig2} superposed on the original 8.3\,\um\, data
(top panel) and on the contrast image (bottom panel).

It is possible that the catalog contains spurious identifications of dark
patches due to holes in the 8 \micron\ emission. Such misidentifications can
only be revealed by a careful cross-correlation of the catalog members with
emission tracing cold molecular gas, namely molecular line or millimeter/submillimeter 
continuum observations. The reliability, including validation of a
sample of clouds in the catalog, is discussed in Section~\ref{reliability}.

\subsection{Cores}

In many cases, individual IRDC candidates show considerable substructure in their contrast images.
Since the distinct mid-IR extinction peaks probably represent column
density peaks, and hence candidate star-forming cores, their positions, sizes,
and contrasts are of potential interest.  Moreover, different dark cores
apparently of the same dark cloud are not necessarily at the same distance, but
may be two distinct cores at different distances along the same line of sight.
This presumption is confirmed for a few high contrast IRDCs from our catalog,
where, for some cores of the same cloud, we found significantly different
molecular line radial velocities (Simon et al., submitted) in \drco\ \jeins\ 
Galactic Ring Survey data \citep[GRS,][]{Simon01}.

To decompose the clouds into cores we use two-dimensional elliptical Gaussian
fits to the contrast images and fit at least one core to each cloud starting
with the peak contrast.  For this first core, the contrast value and the center
position of the Gaussian are fixed to the peak contrast and its position. The
sizes and orientations of the Gaussian ellipses are free fit parameters.

\begin{figure*}[h,t]
\epsscale{0.8}
\plotone{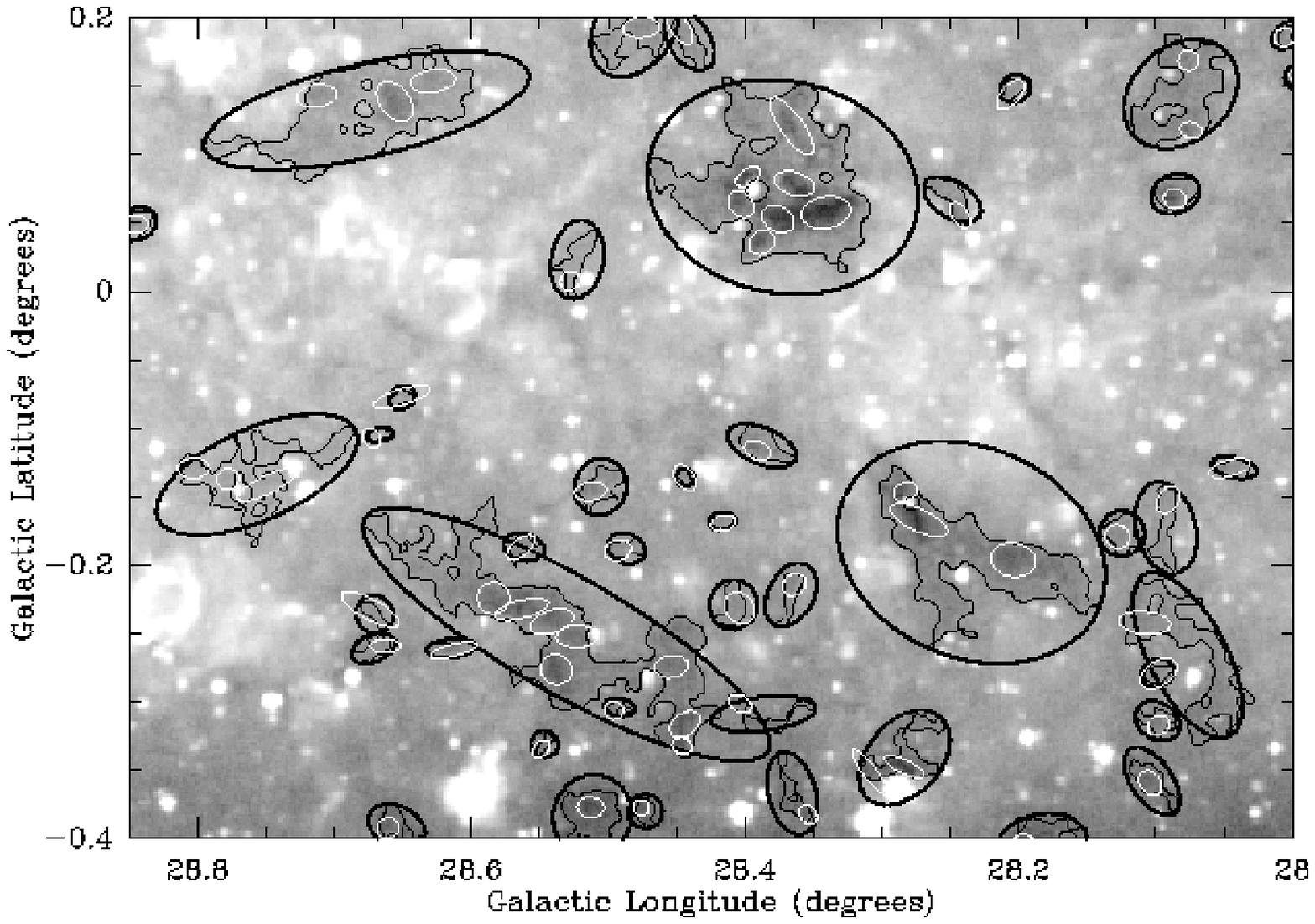}
\plotone{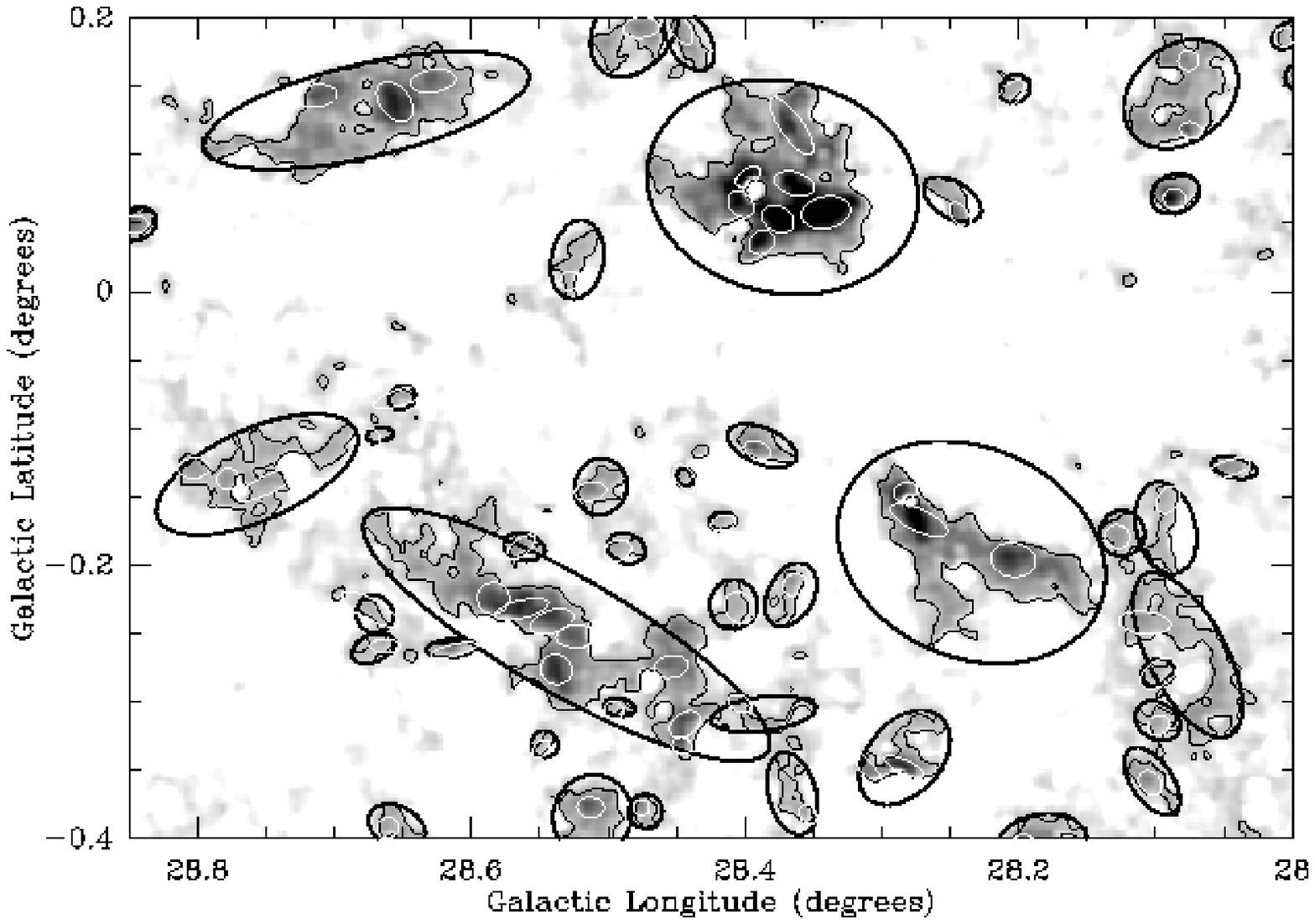}
\caption{Top: \msx\ 8.3 \um\ image of a region along the Galactic plane around
  $l=28$\degree. Clouds identified from our algorithm are marked with thick
  black ellipses, while the cores are marked with white ellipses. The intensity
  scale is logarithmic from $1.5 \times 10^{-6}$ (black) to $1.0 \times 10^{-5}$
  W m$^{-2}$ sr$^{-1}$ (white). The thin contours mark the 0.1 contrast level.
  Bottom: Contrast image with the same contours and ellipses plotted.}
\label{clouds+cores}
\end{figure*}

After subtracting the fit result from the original contrast image, the new peak
contrast and its position in the residual image are determined.  If the residual
peak contrast is at least 40\% larger than the cloud's average contrast
(integrated contrast/area), we fit another core. This procedure was iterated
until the residual peak contrast falls below the 40\% cutoff.  After some
experimentation with the cutoff value, we selected the value of 40\% to find
cores that would be identified by eye, and to exclude fluctuations in the
complicated cloud structure. Fig.~\ref{clouds+cores} also shows the cores
(marked as white ellipses) identified within the clouds for a small region in
the Galactic plane.

The purpose of cataloging the cores is to provide the locations and approximate
extents of the compact highest contrast regions within IRDCs.  Preliminary
results suggest that IRDCs contain real cores that are potentially sites of star
formation \citep{Redman03,Ormel05,Rathborne05}.  Moreover, although the IRDC
position is listed by the contrast centroid, the first core position for each
cloud lists the contrast peak.  To locate the most probable site of current or
future star formation, the position of peak contrast is clearly of interest.
Because the larger high contrast clouds usually have more than one contrast
peak, some sort of decomposition of the contrast distribution
is required in order to identify potentially interesting cores in the catalog.
  
Although the actual cloud and core morphologies are usually complicated,
nevertheless a simple Gaussian decomposition of the highest contrast clump is
usually sufficient to determine positions and approximate sizes. As can be seen
in Fig.~\ref{clouds+cores}, the fitted cores are typically well within the
boundaries of the dark clouds and the positions of the contrast enhancements are
well defined [see, e.g., the IRDC at (l,b)=(28.35,0.05) in the lower panel]. The
orientation and sizes of the cores, however, are sometimes not well fitted, in
particular close to foreground stars visible as holes in the contrast image
[see, e.g., the IRDC at (l,b)=(28.2,--0.2) in Fig.\ref{clouds+cores}].

\subsection{IRDC Reliability}
\label{reliability}

Given the intrinsic difficulties in identifying regions of contrast against a
highly variable background, our identification algorithm will miss some real
IRDCs and also produce some spurious identifications.  The spatial sizes of the
detected IRDCs are further constrained by the finite resolution of \MSX\ and by
the chosen size of our spatial median filter.  The algorithm also cannot
distinguish voids in regions of complicated, bright emission from actual
extinction features. It is also possible that adjacent IRDCs
may be parts of the same object split by foreground emission or unveiled young
stars. Unfortunately, given the very complicated morphologies of
both the sources and the background, the reliability and completeness of the
catalog is difficult to estimate.  The best approach is to measure these
empirically by verifying the reality of a candidate cloud via morphological matching of
molecular line and millimeter/submillimeter continuum emission.  Preliminary studies to do just
that are underway.  In the meantime, we caution the reader that the reality of
specific, individual IRDC candidates will require confirmation.  We assess the
reliability of our catalog in the following sections.

\subsubsection{Size Scales}

\begin{figure*}[h,t]
\epsscale{0.45}
\plotone{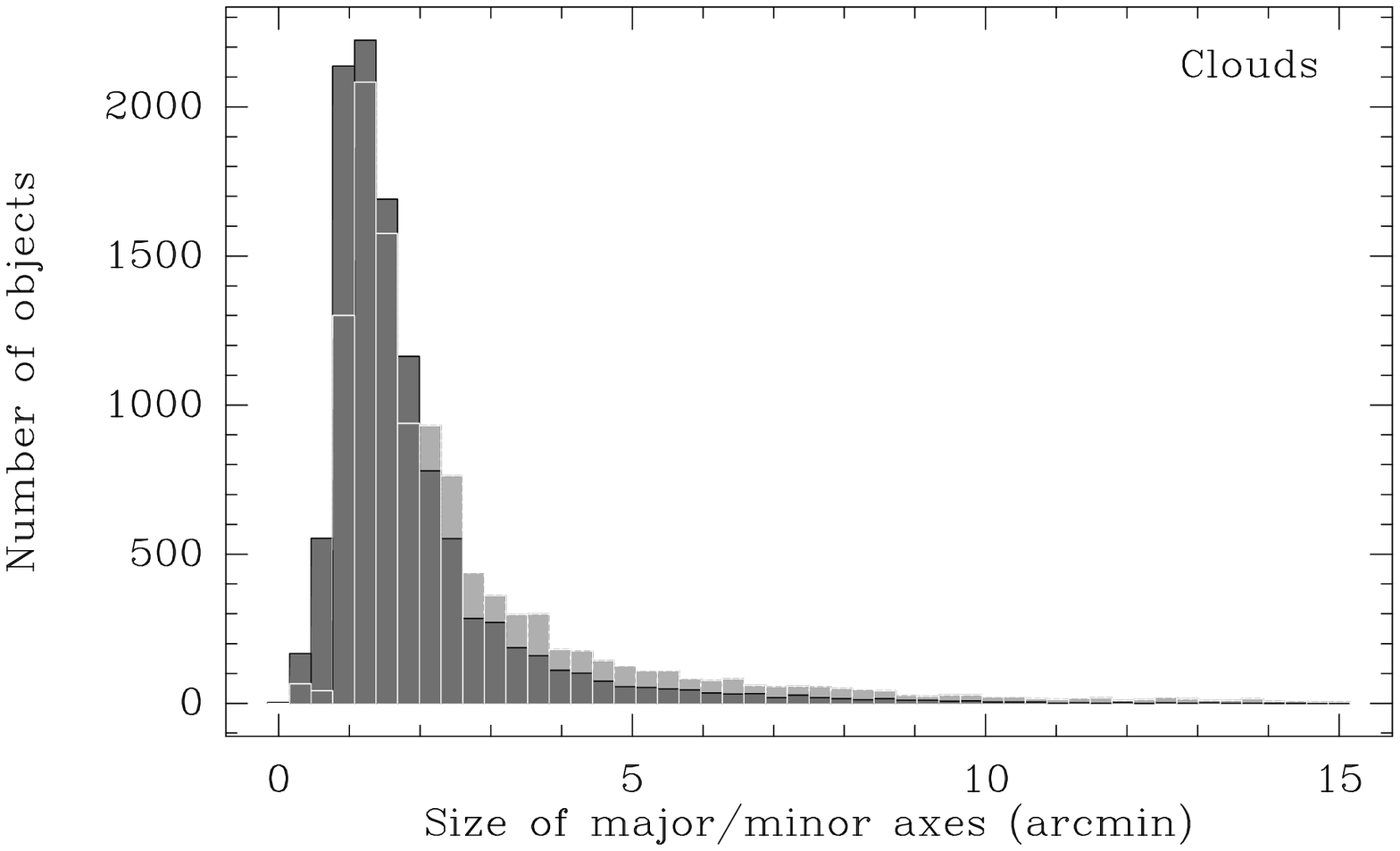} \plotone{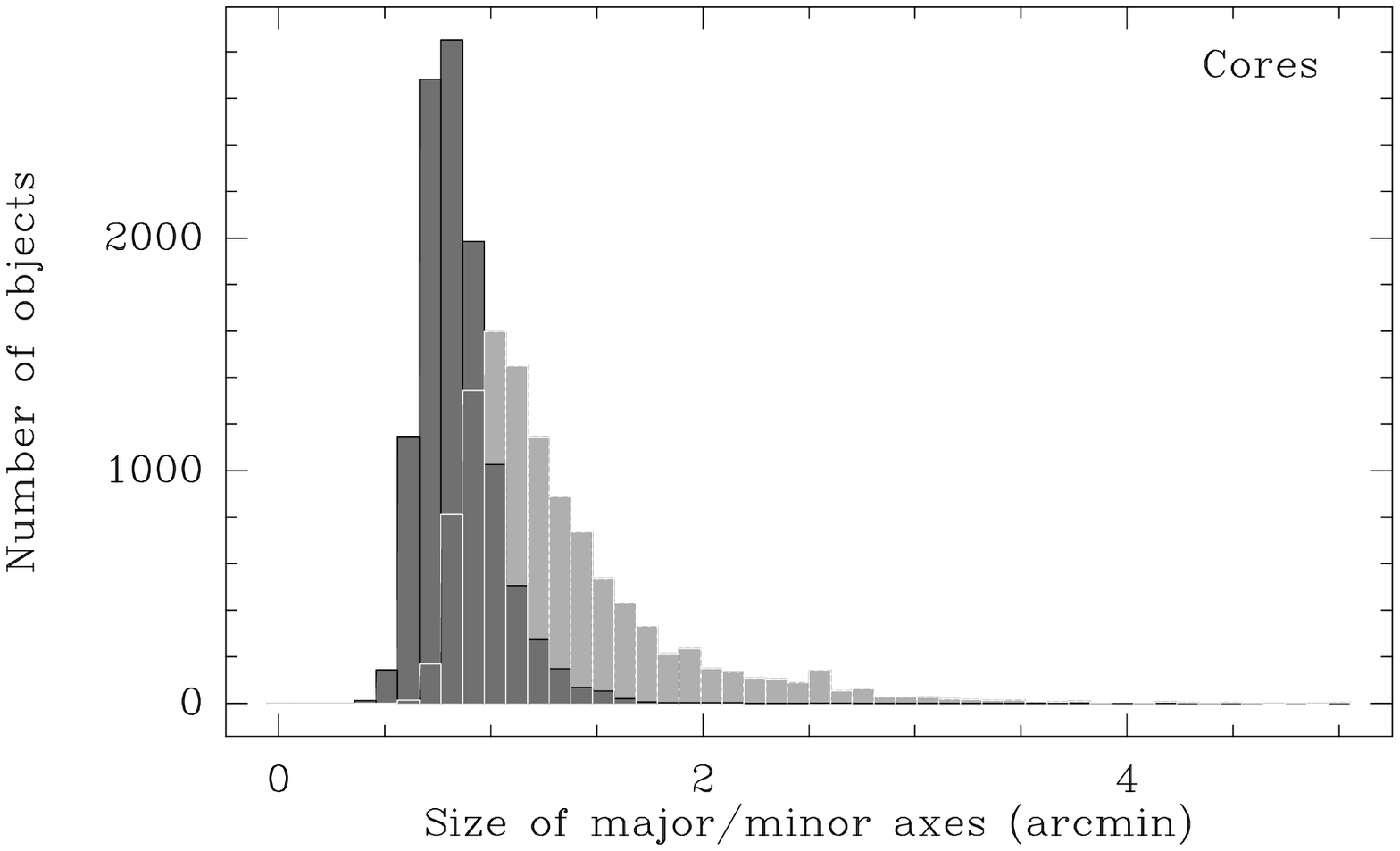}\\
\caption{Number distribution of the major (light gray) and minor (dark gray) axes of the
  clouds (left) and cores (right).}
\label{new1}
\end{figure*}

The minimum and maximum sizes of clouds and cores detectable by our algorithm
are governed by the median filtering circle on the large scales and the angular
resolution in the \msx\ images after smoothing on the small scales. These
angular scales are clearly reflected in the cloud and core statistics shown in
Fig.~\ref{new1}.
  
Typical cloud sizes are 1\arcmin\ to 5\arcmin. We do not detect clouds that
are $>$ 7\farcm5. Although our algorithm may miss clouds $>$ 0\fdg5 (i.e., the diameter of
our filter), visual inspection of the \MSX\ images suggests that no obvious
IRDCs have escaped detection due to their large size.

Typical core sizes are 0\farcm75 to 2\arcmin\ and the minimum detected size
corresponds to the 28\arcsec\ angular resolution of the \msx\ images after
smoothing.

\subsubsection{Source Validation}

To validate a subsample of our IRDCs, we correlated our catalog with the
source lists obtained from \msx\ \citep{Egan98,Carey98,Carey00} and \iso\
data \citep{Hennebelle01,Teyssier02}.

Out of the combined \msx\ lists containing 14 entries, we detect all IRDCs
except the cloud at (l,b)=(79.34,0.33). We do not detect this
cloud as it is located next to the very bright mid-IR emission associated
with the radio continuum source DR15 \citep{Downes1966} and also has two bright,
presumably foreground, compact sources superposed on top. This combination
causes the modelled background towards the dark cloud to be
overestimated, which in turn lowers the contrast below our identification
threshold.

The \iso\ lists of \citet{Hennebelle01} and \citet{Teyssier02} contain 6 and
13 dark cloud candidates (3 identical in both lists), respectively, extracted
using a multi-scale wavelet analysis \citep{Hennebelle01}. We detect all 6
from \citet{Hennebelle01} and 7 out of 13 IRDCs listed in \citet{Teyssier02}.
The six clouds our method misses have either very low contrast in the \msx\
images, are narrow, or small extinction features on top of a very bright
background. We therefore conclude that we do not detect these clouds due to
lower sensitivity, resolution (or a combination of the two), and strong
gradients in the mid-IR emission in complex regions.
 
Another study to test the reliability of our catalog made use of \drco\ 
\jeins\ data from the GRS (Simon et al., submitted). We
identified a sample of 379 large, high-contrast clouds (sizes $>$ 1.53$\arcmin$ 
and contrast $>$ 0.25) from our catalog that fell within the GRS survey region.
If the \drco\ morphology over a particular velocity range matched the mid-IR 
extinction, we concluded that the candidate IRDC is in fact real.  
Out of 379 sources, we established morphological matches with 312 clouds.  
This suggests that, at least for the large, high contrast clouds the reliability 
is close to 82\%. The difficulties in identifying clouds are exacerbated at low contrast
and low signal-to-noise. Therefore, we fully expect the lower contrast clouds to have a lower
reliability. However, quantitatively assessing the reliability of the catalog
as a function of contrast is beyond the scope of the present paper.

\subsubsection{Spurious Identifications in Complex Emission Regions}
\label{complex}

Our algorithm may make spurious IRDC identifications, most often in bright
emission regions with complex structure or at the borders of bright, diffuse
emission. In such regions, almost always star-forming regions such as W51 or
Cygnus X, gaps between bright structures or filaments may erroneously
be identified as IRDC candidates.  Indeed, of the 18\% of the spurious
identifications in the GRS region, most were found toward the bright, complex
W51 region.  The user should be particularly suspicious about the reality of
IRDC candidates toward such regions.
  
\subsubsection{Completeness}

The completeness of the catalog is difficult to estimate. 
The detectability of dark clouds and the observed contrast values depend on
a number of factors, including the instrumental noise level, the background
intensity, and radiative transfer along the line of sight. Toward high latitude
regions the 1$\sigma$ fluctuations, presumably the instrumental noise, is
$2.1 \times 10^{-7}$\,\msxunit. This is an upper limit to the noise as
the \MSX\, Galactic plane survey noise is a function of Galactic longitude
and latitude. The data are slightly more sensitive toward the Galactic plane.

The background varies considerably from
roughly  $1-2 \times 10^{-6}$\,\msxunit in the Galactic Plane to
a maximum of $1.4 \times 10^{-5}$\,\msxunit\, toward the Galactic Center
region. Although in principal we could detect totally opaque clouds with contrast
of 1.0, we do not observe any clouds with such high contrast levels. This
implies that IRDCs are either partially transparent or that the absorption
gets filled in with foreground emission on the way to the observer.
Considering that IRDCs are typically a few kpc away, the latter explanation appears to be 
more likely. Foreground emission would result in a lower observed contrast and may
even cause intrinsically high contrast clouds to drop below our detection
limit. The most straightforward method for finding the completeness and reliability of the
catalog is by empirical studies using molecular line or millimeter/submillimeter continuum
data. 
 
\subsubsection{Contrast Significance}

\begin{figure*}[h,t]
\includegraphics[angle=-90,width=0.95\textwidth]{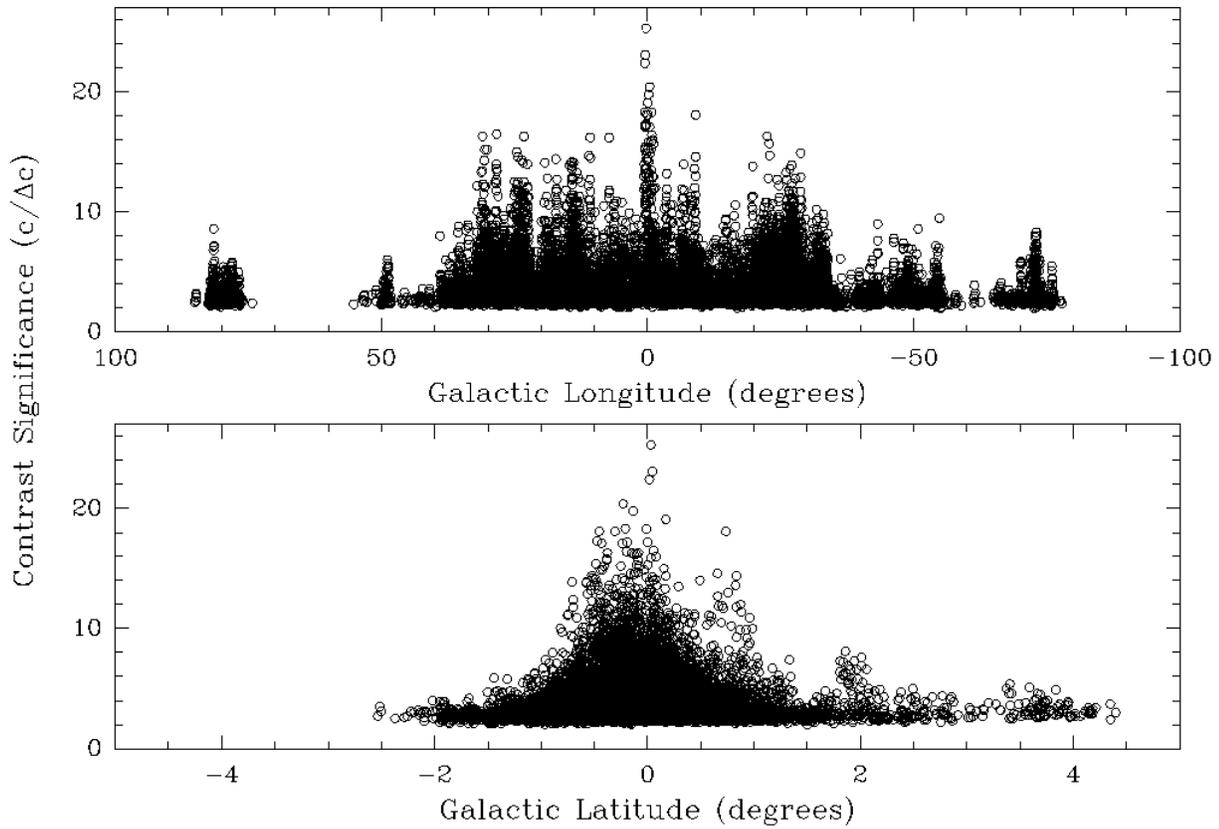}
\caption{Contrast significance as a function of Galactic longitude (top panel) and latitude (bottom panel).}
\label{contrast-sig}
\end{figure*}

To give some estimate of the signal to noise or significance of an IRDC identification, we calculate 
the ratio of the peak contrast to its error (c/$\Delta$c). This was calculated at the center pixel for each
of the cores identified.
We do not calculate this for the clouds, as the position listed in the catalog is the 
contrast centroid and not the position of the peak contrast. 
The contrast error for a particular pixel was calculated by dividing the 
1$\sigma$ noise in the images ($2.1 \times 10^{-7}$\,\msxunit) by the value of the 
modelled background in that pixel. Fig.~\ref{contrast-sig} shows the contrast
significance as a function of Galactic longitude and latitude. 

\section{Description of the Catalog}

The catalog contains entries for each cloud and its associated cores.
Table~\ref{tableone} gives a sample of the catalog entries.  The complete
catalog is available in electronic format at the website
http://www.bu.edu/galacticring/msxirdc.

\subsection{Clouds}

For each IRDC identified, we list several parameters derived directly from the
identification algorithm. For all derived parameters, we consider only pixels
above the $2\sigma$ threshold. The columns of Table~\ref{tableone} are as follows: (1) the
cloud name, designated as MSXDC (for \MSX\, Dark Cloud) followed by the Galactic
longitude and latitude (l and b) coordinates, e.g., MSXDC G045.33+00.65, (2) a descriptor that indicates
whether the entry is a cloud or core.  Clouds are denoted by a ``0'' in this
field, and cores by a letter (a, b, c, etc), (3) and (4) the Galactic
coordinates (l and b) in degrees, calculated as the centroid of the contrast
distribution, (5) the length in arcmin of the major axis of the cloud, defined
as twice the distance between the cloud's contrast centroid and the most distant
pixel from the centroid, (6) the length in arcmin of the cloud's minor axis,
defined as twice the distance between the cloud's contrast centroid and the most
distant pixel from the centroid along the axis perpendicular to the major axis,
(7) the position angle in degrees east of north of the major axis, defined as
the line connecting the cloud's contrast centroid and the cloud's most distant
pixel from the centroid,(8) the total area of the cloud in square arcmin (area
of the pixels above 2$\sigma$), (9) the peak contrast, (10)  the ratio of the core's 
peak contrast to its uncertainty (this is not calculated for clouds), and (11) the integrated
contrast in square arcmin, defined as the sum of the contrast over all pixels
above 2$\sigma$, times the area of the cloud.

Note that the cloud coordinates give the position of the contrast centroid
and do not necessarily reflect the position of the contrast peaks (these are
listed in the information for the cloud cores). As a result, the contrast centroid
may not fall within the cloud boundary for some complex morphologies.
Because the clouds have irregular shapes, we present the results in columns (3) to (7) merely to
estimate a clouds' approximate position, size, and orientation. Because the dark cloud
contrast is roughly proportional to the column density, the integrated
contrast given in column (11), just like the integrated intensity of an
optically thin molecular line map, should be roughly proportional to the mass
of the cloud.

\subsection{Cores}

The cores, or contrast peaks, were found by fitting elliptical Gaussians to the
contrast image for each cloud.  Every cloud has at least one core fitted at the
peak contrast position, and larger clouds may contain additional cores.
In the catalog, the cores are listed immediately after the entry for their associated
cloud.  The columns are as follows:
(1) the name of the IRDC that contains the core, 
designated as  MSXDC (for \MSX\, Dark Cloud) followed by the Galactic coordinates,
e.g., MSXDC G045.33+00.65,
(2) a descriptor that indicates whether the entry is a cloud or core.
Cores are designated by a letter (a, b, c, etc) in order of  decreasing contrast,
(3) and (4) the Galactic coordinates (l and b) in degrees of the core position, defined as the 
peak of the fitted elliptical Gaussian, 
(5) the length in arcmin of the major axis of the fitted ellipse,
(6) the length in arcmin of the minor axis of the fitted ellipse,
(7) the position angle in degrees east of north of the major axis of the fitted ellipse, 
(8) the total area of the core in square arcmin, 
(9) the peak contrast, 
(10) the ratio of the peak contrast to its uncertainty, and
(11) the integrated contrast in square arcmin, defined as the integrated
contrast of the elliptical Gaussian within the half power contour.

\section{Ensemble Properties}

\begin{figure*}[h,t]
\epsscale{1.0}
\plotone{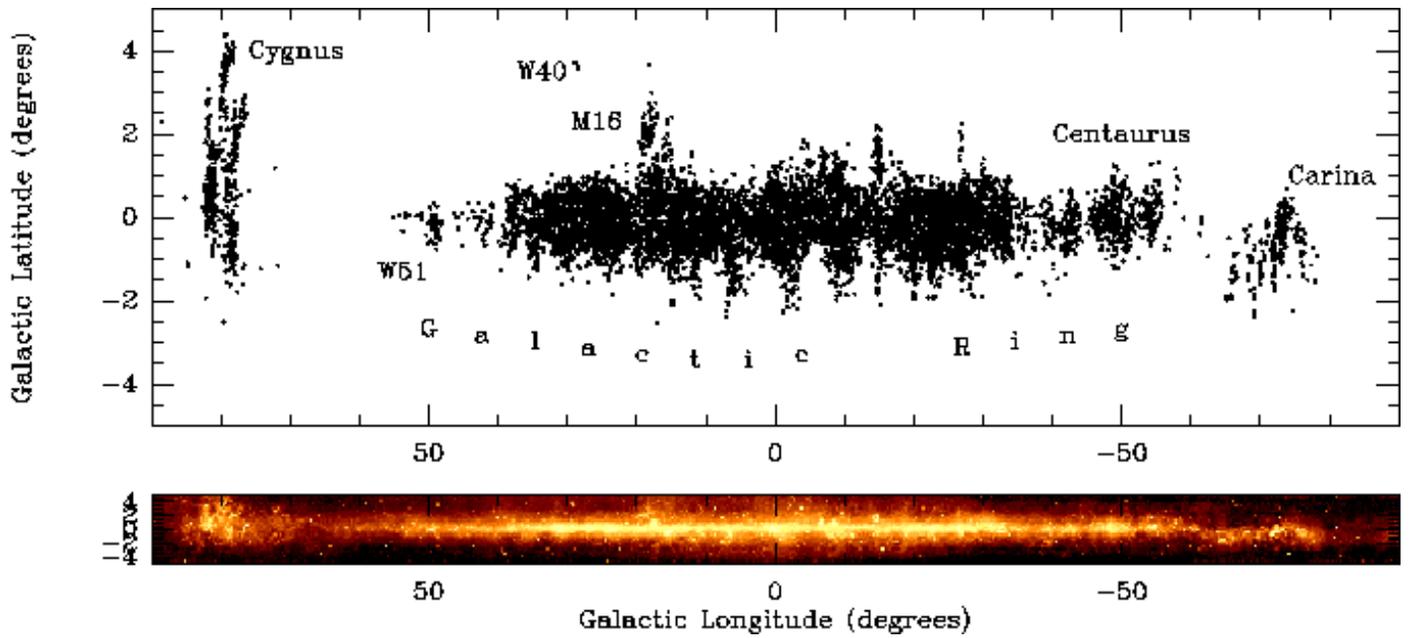}
\caption{Distribution of IRDCs along the first and fourth quadrant of the Galactic
  plane. Prominent star forming regions are labeled. The latitude axis of the
  plot is stretched in the top panel.  For comparison, \msx\ emission at 8.3 \micron\ from the
  same region is shown in the bottom panel.}
\label{irdc-distr}
\end{figure*}

The Galactic distribution of the IRDCs is shown in Fig.~\ref{irdc-distr}.
Not surprisingly, the IRDC distribution is well correlated with bright 8.3\,\micron\ 
emission from the Galactic plane. Because IRDCs are extinction
features, they are easier to detect when the background emission is bright. In
addition, since IRDCs may be associated with star forming regions,
their number may be enhanced where the background is brightest. With the
current data, it is difficult to distinguish between these two factors.
  
In detail, IRDCs are more numerous toward prominent IR-bright, star forming
regions in the inner Galaxy such as the spiral arm tangents toward Cygnus
(Orion arm) and W51 (Sagittarius arm) in the north, and Centaurus (Norma arm)
and Carina (Scutum arm) in the south. IRDCs are also common toward the bright
IR emission between Galactic longitudes $-50^\circ$ to $+50^\circ$ due to
heated dust associated with the Galactic 5 kpc molecular Ring, the Milky Way's
dominant star forming structure \citep{Robinson84}.

\begin{figure*}[h,t]
 \plotone{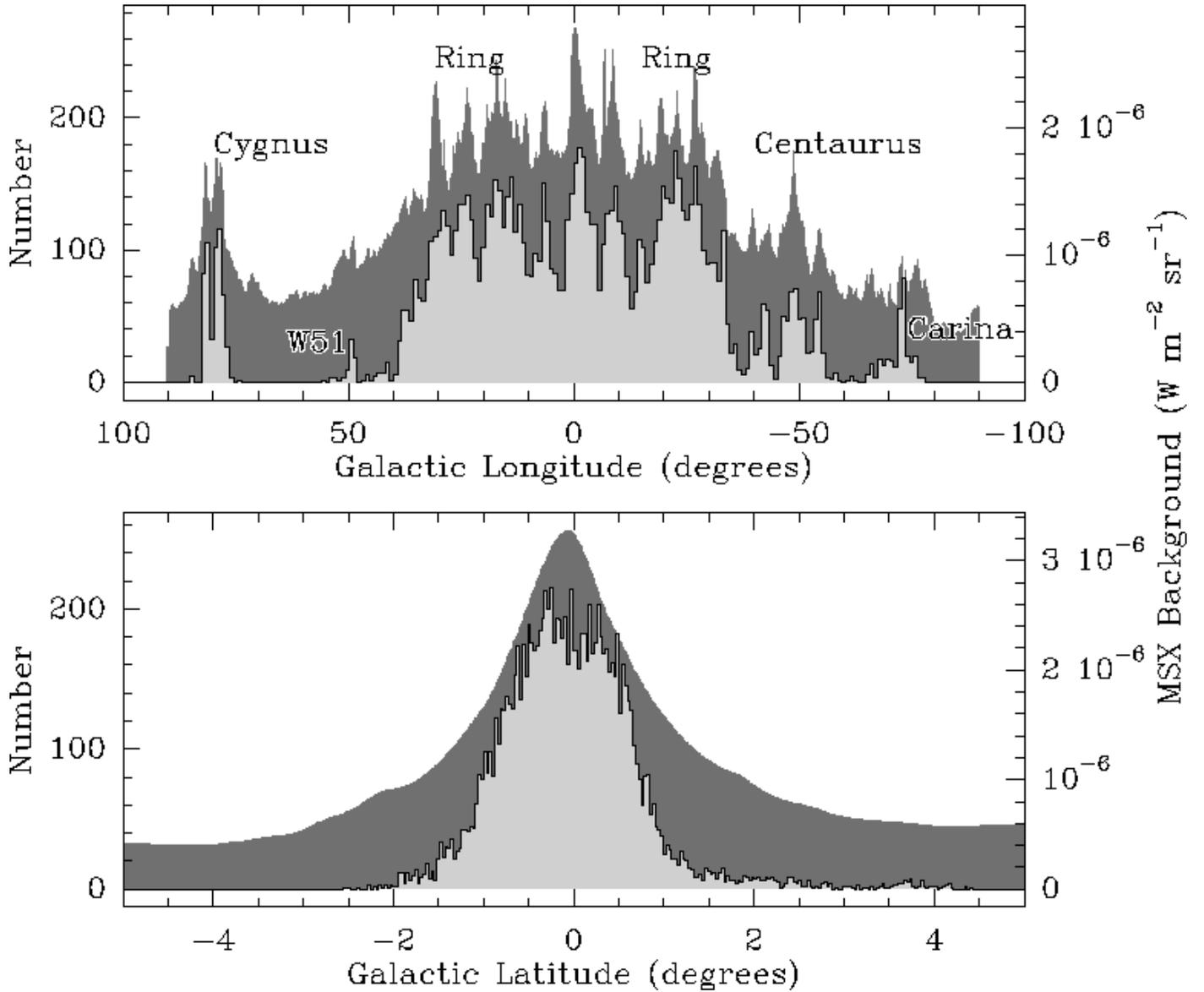}
\caption{Number distribution (light gray histograms) of IRDCs with Galactic longitude
  (top) and latitude (bottom) against \msx\ background emission (dark gray
    histograms, right axes) modelled as described in the text and averaged over
    latitude and longitude, respectively. Prominent Galactic star forming
  regions and the longitude range of the Galactic Ring are labeled in the top
  panel.}
\label{histo-lb}
\end{figure*}

\begin{figure}[h,t]
\plotone{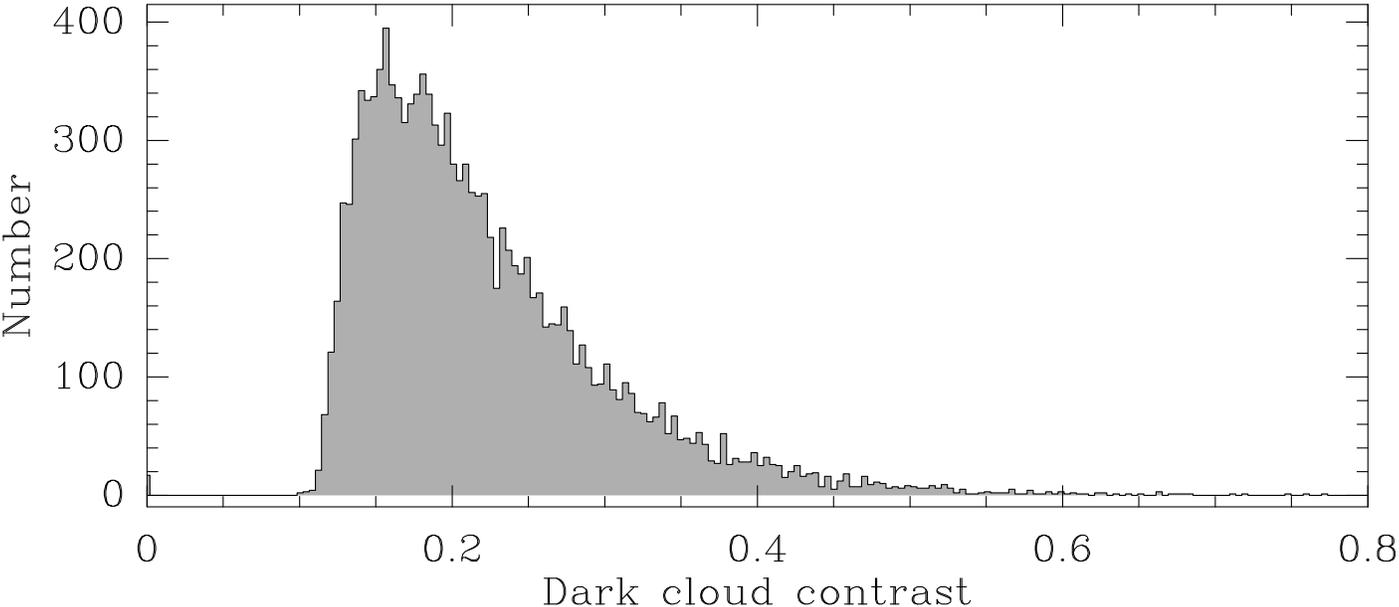}
\caption{Number distribution of IRDCs with peak contrast. The distribution is approximately
power law and sharply declines with increasing contrast.}
\label{histo-contrast}
\end{figure}

In Fig.~\ref{histo-lb} we plot histograms of the IRDC distribution as functions
of Galactic longitude and latitude together with the variation in the modelled
background intensity.  Both distributions reflect the diffuse mid-IR background.
The latitude distribution is approximately Gaussian with a FWHM of $\sim
2$\fdg2.  Very few IRDCs are found beyond $\pm 2^\circ$ in Galactic latitude.
In Fig.~\ref{histo-contrast} we plot a histogram of the IRDC peak contrast.  The
distribution is approximately power law and sharply
declines with increasing contrast.  The turnover near a contrast of 0.1 reflects
the sensitivity limit of the \msx\ data and our identification algorithm.

\section{Summary}

We present a catalog of candidate IRDCs in the Galactic plane spanning the range of
Galactic longitudes $-90^\circ$ to $+90^\circ$ and latitudes $-5^\circ$ to
$+5^\circ$.  We identified the clouds using algorithms which model the diffuse
IR background via a median filter technique, generate images of decremental
contrasts, and select IRDCs based on contiguous pixels above a $2\sigma$
contrast threshold.  A further criterion is that the IRDCs must be extended, so
that they cover a solid angle at least twice that of the 28$''$ convolved beam
solid angle.  We have identified a total of 10,931 IRDCs.  

Within each cloud, we have also identified embedded cores, or contrast peaks.  For each cloud, we
identify a core with the maximum contrast peak, and also include other cores if
their contrasts are $>40$\% of the IRDC's average contrast.  We identify 12,774
cores.  

The Galactic distribution of IRDCs closely follows the mid-IR background
emission, and the distribution of contrasts is approximately power-law.  In the
future, radio, millimeter, submmillimeter, and IR studies will yield the distances, sizes,
masses, and other physical properties of IRDCs and their embedded cores, and
establish their role in the process of star formation.

\acknowledgements {We thank the referee Sean Carey for his comments and
    suggestions to improve the paper. This research made use of data products
  from the {\it Midcourse Space Experiment}.  Processing of the data was funded by the
  Ballistic Missile Defense Organization with additional support from NASA
  Office of Space Science. This research has also made use of the NASA/ IPAC
  Infrared Science Archive, which is operated by the Jet Propulsion Laboratory,
  California Institute of Technology, under contract with the National
  Aeronautics and Space Administration. This research has made use of NASA's
  Astrophysics Data System Abstract Service and was partially funded by NASA
  under grants NNG04GC92G and NAG5-10808.}

\begin{table*}[t,h]
\scriptsize
\caption{A sample of the \MSX\, IRDC Candidate catalog} 
\label{tableone}
\begin{tabular}{ccccccrccccc}\hline\hline
Cloud name & & \multicolumn{2}{c}{Coordinates} & Major & Minor & \multicolumn{1}{c}{PA} & Area & \multicolumn{3}{c}{Contrast}\\
           & & {\it {l}}  &  {\it {b}} &  &  &  &  & Peak & c/$\Delta$c & Integrated\\
           & & (\degree) &  (\degree) & (\arcmin) & (\arcmin) &  \multicolumn{1}{c}{(\degree)} & (arcmin$^{2}$) &  & & (arcmin$^{2}$) \\
(1) & (2) & (3) & (4) & (5) & (6) & (7) & (8) & (9) & (10) &(11)\\ 
\hline

MSXDC G000.00+00.65   &(0)   & -0.005 &  0.659  &  2.9 &   2.0 &  155  &  1.92 &   0.20  &  ...  &  0.27\\
MSXDC G000.00+00.65   &(a)   & -0.001 &  0.654  &  0.9 &   0.8 &  125  &  0.54 &   0.20  &  4.0  &  0.08\\
MSXDC G000.00$-$00.59 &(0)   & -0.001 & -0.600  &  1.3 &   1.3 &  161  &  0.60 &   0.15  &  ...  &  0.07\\
MSXDC G000.00$-$00.59 &(a)   & -0.001 & -0.602  &  1.1 &   0.7 &  125  &  0.62 &   0.15  &  3.7  &  0.06\\
MSXDC G000.00$-$00.74 &(0)   &  0.001 & -0.744  &  0.9 &   0.9 &  153  &  0.37 &   0.16  &  ...  &  0.05\\
MSXDC G000.00$-$00.74 &(a)   &  0.001 & -0.744  &  1.3 &   0.6 &   95  &  0.61 &   0.16  &  3.6  &  0.07\\
MSXDC G000.00$-$00.33 &(0)   &  0.004 & -0.332  &  4.0 &   3.4 &   36  &  3.73 &   0.18  &  ...  &  0.48\\
MSXDC G000.00$-$00.33 &(a)   &  0.008 & -0.319  &  1.3 &   1.0 &   99  &  1.00 &   0.18  &  9.0  &  0.13\\
MSXDC G000.00$-$00.47 &(0)   &  0.004 & -0.476  &  3.6 &   1.5 &  128  &  1.79 &   0.25  &  ...  &  0.27\\
MSXDC G000.00$-$00.47 &(a)   & -0.001 & -0.476  &  1.6 &   0.7 &  146  &  0.87 &   0.25  &  8.9  &  0.15\\
MSXDC G000.00$-$00.98 &(0)   &  0.005 & -0.989  &  1.3 &   1.3 &   18  &  0.45 &   0.20  &  ...  &  0.08\\
MSXDC G000.00$-$00.98 &(a)   &  0.001 & -0.996  &  0.8 &   0.4 &  215  &  0.28 &   0.20  &  2.4  &  0.04\\
MSXDC G000.00$-$00.78 &(0)   &  0.006 & -0.784  &  1.0 &   1.0 &   53  &  0.49 &   0.18  &  ...  &  0.07\\
MSXDC G000.00$-$00.78 &(a)   &  0.006 & -0.786  &  0.9 &   0.8 &  218  &  0.53 &   0.18  &  3.7  &  0.07\\
MSXDC G000.01+00.63   &(0)   &  0.015 &  0.636  &  1.0 &   1.0 &  168  &  0.67 &   0.13  &  ...  &  0.08\\
MSXDC G000.01+00.63   &(a)   &  0.018 &  0.634  &  1.2 &   1.1 &  123  &  1.09 &   0.13  &  2.7  &  0.10\\
MSXDC G000.02+00.57   &(0)   &  0.021 &  0.577  &  3.8 &   3.8 &   47  &  4.71 &   0.16  &  ...  &  0.61\\
MSXDC G000.02+00.57   &(a)   &  0.028 &  0.581  &  2.2 &   0.8 &  104  &  1.38 &   0.16  &  3.5  &  0.16\\
MSXDC G000.02$-$00.98 &(0)   &  0.024 & -0.983  &  1.1 &   1.1 &   21  &  0.50 &   0.23  &  ...  &  0.10\\
MSXDC G000.02$-$00.98 &(a)   &  0.023 & -0.982  &  1.3 &   0.7 &  235  &  0.70 &   0.23  &  2.8  &  0.11\\
MSXDC G000.03$-$00.35 &(0)   &  0.037 & -0.352  &  1.6 &   0.7 &  104  &  1.00 &   0.14  &  ...  &  0.12\\
MSXDC G000.03$-$00.35 &(a)   &  0.033 & -0.352  &  1.1 &   1.1 &  245  &  0.94 &   0.14  &  6.8  &  0.09\\
MSXDC G000.04$-$00.64 &(0)   &  0.048 & -0.645  &  1.3 &   0.7 &  108  &  0.50 &   0.14  &  ...  &  0.06\\
MSXDC G000.04$-$00.64 &(a)   &  0.053 & -0.646  &  1.3 &   0.6 &  117  &  0.63 &   0.14  &  3.3  &  0.06\\
MSXDC G000.05$-$00.55 &(0)   &  0.057 & -0.555  &  0.9 &   0.8 &   63  &  0.43 &   0.15  &  ...  &  0.05\\
MSXDC G000.05$-$00.55 &(a)   &  0.056 & -0.556  &  1.3 &   0.8 &  241  &  0.87 &   0.15  &  3.7  &  0.09\\
MSXDC G000.05$-$00.68 &(0)   &  0.059 & -0.688  &  5.2 &   2.2 &  125  &  6.48 &   0.29  &  ...  &  1.02\\
MSXDC G000.05$-$00.68 &(a)   &  0.056 & -0.696  &  1.1 &   0.8 &  125  &  0.71 &   0.29  &  6.8  &  0.15\\
MSXDC G000.05$-$00.68 &(b)   &  0.076 & -0.701  &  1.9 &   0.9 &   95  &  1.38 &   0.22  &  5.2  &  0.21\\
MSXDC G000.06+00.21   &(0)   &  0.060 &  0.216  &  3.2 &   1.8 &   71  &  2.32 &   0.22  &  ...  &  0.34\\
MSXDC G000.06+00.21   &(a)   &  0.056 &  0.213  &  1.1 &   0.8 &  246  &  0.70 &   0.22  &  7.9  &  0.11\\
MSXDC G000.06$-$00.77 &(0)   &  0.065 & -0.779  &  8.0 &   6.6 &  121  & 17.74 &   0.24  &  ...  &  2.61\\
MSXDC G000.06$-$00.77 &(a)   &  0.101 & -0.782  &  1.5 &   0.9 &  217  &  1.08 &   0.24  &  5.4  &  0.18\\
MSXDC G000.06$-$00.77 &(b)   &  0.074 & -0.772  &  1.5 &   1.0 &  260  &  1.17 &   0.21  &  4.7  &  0.17\\
MSXDC G000.07+00.24   &(0)   &  0.077 &  0.243  &  1.3 &   0.8 &  108  &  0.62 &   0.16  &  ...  &  0.08\\
MSXDC G000.07+00.24   &(a)   &  0.074 &  0.244  &  1.0 &   0.7 &  185  &  0.50 &   0.16  &  5.6  &  0.06\\
\hline
\end{tabular}
\tablecomments{The complete version of this table is in the 
electronic edition of the Journal and is also accessible at the website http://www.bu.edu/galacticring/msxirdc.}
\end{table*}

\end{document}